# Electromagnetically Induced Distortion of a Fibrin Matrix with Embedded Microparticles

Tyler Scogin[a], Sumith Yesudasan[b], Mitchell L.R. Walker[a], Rodney D. Averett[b*]

[a]College of Engineering, Georgia Institute of Technology, Atlanta, GA, USA
[b]College of Engineering, The University of Georgia, Athens, GA, USA

*Corresponding author email: raverett@uga.edu

**ABSTRACT**

Blood clots occur in the human body when they are required to prevent bleeding. In pathological states such as diabetes and sickle cell disease, blood clots can also form undesirably due to hypercoagulable plasma conditions. With the continued effort in developing fibrin therapies for potential life-saving solutions, more mechanical modeling is needed to understand the properties of fibrin structures with inclusions. In this study, a fibrin matrix embedded with magnetic micro particles (MMPs) was subjected to a magnetic field to determine the magnitude of the required force to create plastic deformation within the fibrin clot. Using finite element (FE) analysis, we estimated the magnetic force from an electromagnet at a sample space located approximately 3 cm away from the coil center. This electromagnetic force coupled with gravity was applied on a fibrin mechanical system with MMPs to calculate the stresses and displacements. Using appropriate coil parameters, it was determined that application of a magnetic field of 730 A/m on the fibrin surface was necessary to achieve an electromagnetic force of 36 nN (to engender plastic deformation).

*Keywords*: fibrin matrix; mechanical behavior; electromagnetic field; magnetic micro-particles; finite element analysis

## 1. Introduction

Fibrin clots are extracellular matrix structures that are critical for the prevention of blood loss during injury to the endothelial lining. Fibrin is the resulting product formed from the soluble molecule fibrinogen when the serine protease thrombin cleaves fibrinopeptides from the molecule, and causes a spontaneous creation of fibrin fibers. These fibrin fibers are the result of fibrin monomers that assemble longitudinally and laterally [1, 2]. The fibers eventually form a meshed network that has a primary purpose of preventing further blood loss. Many experimental studies have been conducted to determine the mechanical behavior of native fibrin clots with and without cells [3-8]. Studying the mechanical behavior of fibrin clots with inclusions (cells or other particles) is important because it can lead to the discovery of therapies and strategies for understanding potential life-threatening clot structures. Thus more mechanical modeling studies need to be conducted on fibrin matrices with the inclusion of other particles, such as magnetic micro-particles (MMPs), since they can be used to ascertain how these forces may exert forces and distort a fibrin matrix when perturbed by a magnetic field.

In a previous experimental study, fibrin networks with embedded quantum dots (QDs) and MMPs were distorted due to the application of an external magnetic field subsequent to polymerization [9]. The MMPs influenced by the magnetic field caused the overall structure of the fibrin matrix to be more anisotropic, so that the fibrin fibers were mostly aligned in one direction. In addition, rheology experiments were used to determine the mechanical properties of the fibrin gels, and it was concluded that noticeable plastic deformation occurred in the samples at 0.1 wt % of content for both the QDs and MMPs, under 10 Hz oscillatory conditions. In the past, researchers researchers have experimentally investigated the polymerization of fibrin under the influence of magnetic fields with the absence of inclusions [10-13]. These studies all show that fibrin becomes highly oriented when polymerized under high magnetic fields.

The primary objective of the current study was to perform finite element simulations of the fibrin matrix with embedded MMPs while subjected to an external magnetic field. These simulations serve as a first-order approximation of the strength of the magnetic field that can be used to distort a fibrin matrix, based on number of turns in the solenoid, size of the solenoid core, and distance from the solenoid to the sample. Input parameters to the FE model include the material properties of the fibrin network sample (i.e. elastic modulus, Poisson's ratio, density of the matrix, electrical conductivity, relative permittivity, and relative magnetic permeability). These input parameters were employed with the finite element analysis to simulate the magnitude and distribution of stresses the MMPs exerted upon the surrounding fibrin network, ultimately creating a strain in the system. For the solenoid-sample area model, the simulation demonstrated the magnitude and direction of the magnetic force necessary to distort the fibrin sample. This mechanical modeling study can be used to aid in the design of experimental or clinical applications that aid with thrombolysis, where an electromagnetic field could be used to dissolve a fibrin clot.

## 2. Methods
### 2.1. Simulation Methodology
An electromagnetic configuration (solenoid) was designed using SolidWorks [14]. Figure 1 displays the configuration in relation to the fibrin matrix sample, as well as its relative position in the confocal microscope platform. The SolidWorks drawing of the microscope is based on measurements taken of the actual Zeiss 700 B confocal microscope located at the Georgia Institute of Technology, which is the same confocal microscope used in the previous experimental fibrin/MMP study [9]. The microscope uses a platform that has adjustable sliders to capture the sample slide, which allow the solenoids to maneuver vertically in the open space. The properties of the electromagnet can be altered, such as the number of turns, electrical current, and composition of wire.

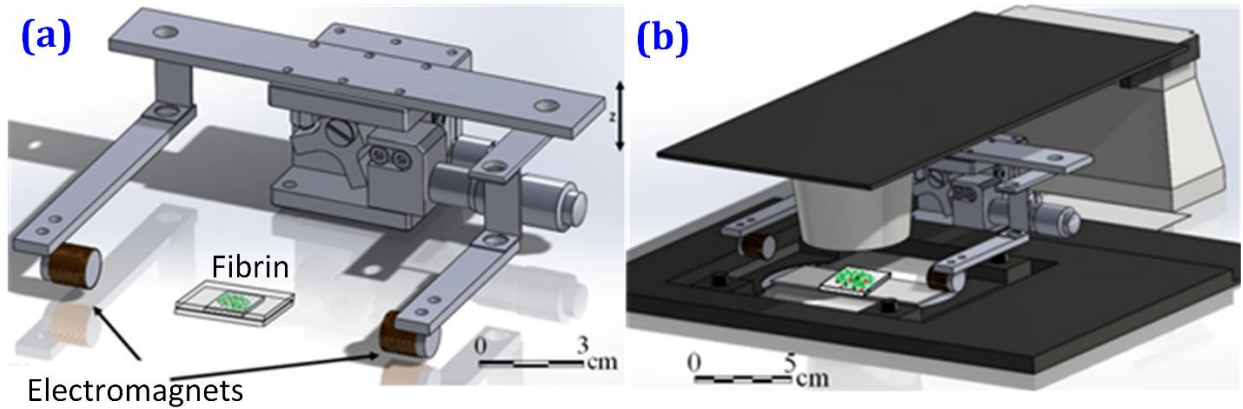

Figure 1: (a) Schematic of mount with electromagnets (solenoids) and (b) electromagnets positioned in confocal microscope

To study the system as shown in Figure 1 using finite element simulations, we adopted a similar method to multigrid analysis [15]. Firstly, we modeled and analyzed the current carrying wires (electromagnets) and the surrounding space to estimate the detailed information of the magnetic field, magnetic vector potential, and magnetic flux density surrounding the electromagnet (Case 1). With information on the tentative location of the fibrin clot sample, we estimated the magnetic field, magnetic vector potential, and magnetic flux acting on the fibrin sample from Case 1. This information was utilized as initial and boundary conditions to a sub-model of fibrin with dispersed MMPs. For the latter case (Case 2), a magnetic analysis was performed and the Maxwell forces on the MMPs were calculated. The calculated forces were used as inputs for the elastic equation to solve for stresses and displacements.

### 2.2. Mathematical Formulation
The forces acting on an MMP embedded in a fibrin matrix were modeled using FE analysis. The magnetic forces are balanced by the combination of elastic forces and gravity. The resultant net force was used to determine if the system resided in static or dynamic equilibrium.

The equations pertaining to the force equilibrium are:

$$\sum \vec{F}_X = m\vec{a}_x = \vec{F}_{x,magnetic} - \vec{F}_{x,elastic} \quad (1)$$

$$\sum \vec{F}_Y = m\vec{a}_y = \vec{F}_{y,magnetic} - \vec{F}_{y,elastic} \quad (2)$$

$$\sum \vec{F}_Z = m\vec{a}_z = -\vec{F}_{z,magnetic} + \vec{F}_{z,elastic} - \vec{F}_{gravity} \quad (3)$$

In (1) – (3), *m* is the mass of the MMP (diameter = 2.8 μm, density = 7,800 kg/m$^3$), and the subscripts *magnetic*, *elastic*, and *gravity* represent the magnetic forces from the current carrying coil, elastic resistance force from the fibrin, and gravitational force, respectively. The equations related to the magnetic field induced due to the current carrying coil are prescribed in (Eq. 4-8). Equation 4, Ampere's law, estimates the magnetic field $\overline{H}$ and the magnetic flux density $\overline{B}$ surrounding a current carrying conductor with current density $\overline{J}_e$. $\overline{v}$ represents the velocity of the moving charge (in the present work $\overline{v}=0$).

$$\nabla \times \overline{H} - \sigma \overline{v} \times \overline{B} = \overline{J}_e \qquad (4)$$
$$\overline{B} = \nabla \times \overline{A} \qquad (5)$$
$$\overline{J}_e = \nabla \times \overline{H} \qquad (6)$$
$$\overline{B} = \mu_0 \mu_r \overline{H} \qquad (7)$$
$$\overline{D} = \varepsilon_0 \varepsilon_r \overline{E} \qquad (8)$$

In (Eq. 4-8) $\sigma$ is the electrical conductivity, $\overline{A}$ is the magnetic vector potential, $\varepsilon_0$ is the permittivity of vacuum (8.8542 x 10$^{-12}$ *F/m*), $\mu_0$ is the permeability of vacuum ($4\pi 10^{-7}$ *N/A$^2$*), $\mu_r$ and $\varepsilon_r$ are the relative permeability and permittivity of the material, respectively, $\overline{E}$ is the electric field, and $\overline{D}$ is the electric flux density.

Utilizing the magnetic field and flux density, the electromagnetic forces acting on a material using Maxwell's stress tensor were calculated using (9). For the current study, under vacuum conditions in the absence of moving charges (9) reduces to (10).

Air:
$$\overline{\overline{T}} = -p\overline{\overline{I}} - \left(\tfrac{1}{2}\overline{E}\cdot\overline{D} + \tfrac{1}{2}\overline{H}\cdot\overline{B}\right)\overline{\overline{I}} + \overline{E}\,\overline{D}^T + \overline{H}\,\overline{B}^T + (\overline{D} \times \overline{B})\overline{v}^T \qquad (9)$$
Vacuum:
$$\overline{\overline{T}} = -\left(\tfrac{1}{2}\overline{E}\cdot\overline{D} + \tfrac{1}{2}\overline{H}\cdot\overline{B}\right)\overline{\overline{I}} + \overline{E}\,\overline{D}^T + \overline{H}\,\overline{B}^T \qquad (10)$$

In (9) – (10), $\overline{\overline{T}}$ is Maxwell's stress tensor, *p* is the air pressure, and $\overline{\overline{I}}$ is the identity matrix. The Maxwell stress tensor was integrated over the surface area to obtain the corresponding force using (Eq. 11).

$$\overline{F}_{EM} = \int_{\partial\Omega} \overline{n}\cdot\overline{\overline{T}}\,ds \qquad (11)$$

In (11), $\overline{F}_{EM}$ is the electromagnetic force, $\overline{n}$ is the unit normal vector to the surface, and $\overline{\overline{T}}$ is the Maxwell tensor. The electromagnetic force was used to calculate the displacement and stress solution by coupling with the solid mechanics equation as provided in (12).

$$\rho \frac{\partial^2 \overline{u}}{\partial t^2} = \nabla \cdot \overline{\overline{S}} + \overline{F} \qquad (12)$$

In (12), $\overline{\overline{S}}$ is the stress tensor, $\overline{F}$ is the external force, $\rho$ is the density of the material, and $\overline{u}$ is the displacement.

$$\overline{F} = \overline{F}_{EM} + m\overline{g} \qquad (13)$$

In (Eq. 13), *m* represents the mass and $\overline{g}$ is gravity (z-axis).

## 3. Finite Element Simulations

The finite element simulations were approached in a sequential manner. The simulations were performed using the COMSOL Multiphysics® (Stockholm, Sweden) FE package [16]. The first simulation (Case 1) with the AC/DC module consisted of an arrangement (Figure 2a). The surface force on this solid was assumed to be surrounded by a vacuum (Eq. 10). This was approximated in the simulation by replacing the vacuum with air (Eq. 9) when calculating forces on the surface of the solid. The domains adjacent to the model that were approximated by vacuum or air were assumed to have a negligible influence on the electromagnetic fields. This analysis was followed by a sub model analysis and will be discussed in the subsequent section.

### 3.1. Case 1: Magnetic Field and Flux Estimation

The objective of this case was the simulation of magnetic field and estimation of magnetic flux and potential due to a current carrying coil. The coil represents the solenoid (Figure 1) and the sample space represents the fibrin clot sample. The coil was modeled as circular rings with a diameter of 1 cm and a wire cross sectional area of 0.01 cm$^2$. There are 10 coils with 19 turns in each coil and each turn carries 1 A of current. The sample space, which represents the actual location of the fibrin clot, was modeled as a cube with dimensions of 20 μm x 20 μm x 20 μm at a distance of 3 cm from the coil. Figure 2(b) shows the meshed coil and sample surrounded by air. The enlarged view of both the coil and sample space with mesh and without mesh is shown in Figure 2d and Figure 2c, respectively. Further the coil and sample space is enlarged with meshing details (Figure 2f, Figure 2g). Second-order tetrahedral elements were used for the analysis. The level of mesh refinement adopted in FE simulations was determined via convergence analysis. Figure 3e shows the naming convention used to identify the faces of the sample space. The first letter indicates the direction of the normal to the surface, and low and high are binary descriptions used to indicate the location of the face (near origin-low, away from origin-high). This naming scheme was used to store the results of the magnetic simulations from Case 1. The space surrounding the system was air and modeled as a cylinder with a height of 10 cm and radius of 10 cm.

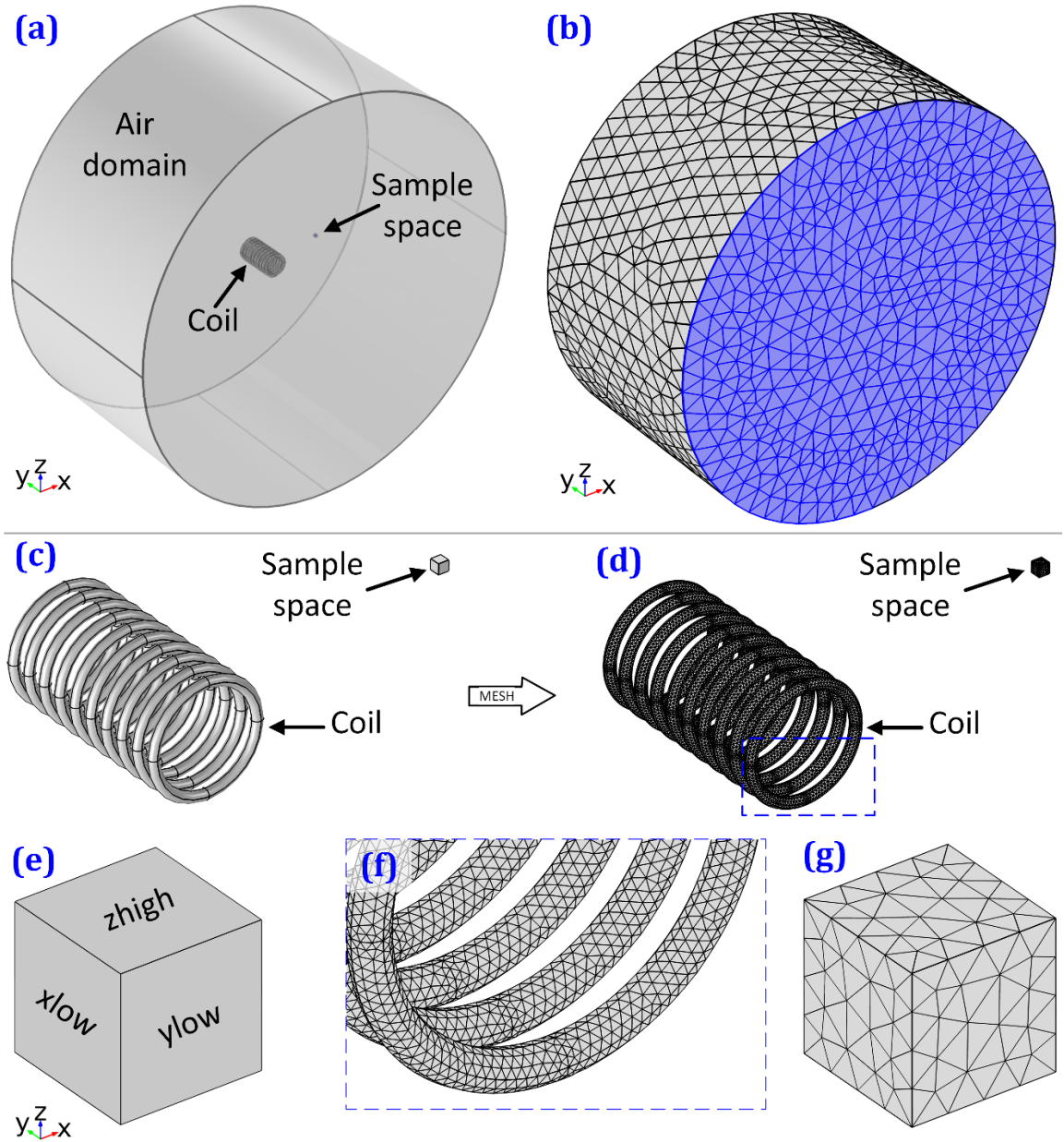

**Figure 2**: Magnetic flux density and magnetic field estimation. (a) 3D model of the coil, fibrin sample space and the surrounding medium (air), (b) Meshed model for the system, (c) zoomed view of the coil and the sample, (d) zoomed view of the meshes of coil and sample space, (e) naming of the sample space surfaces. (f) zoomed-in details of coil mesh. (g) Mesh of the sample space

The electromagnetic equations (4) - (8) were then solved, which provided the magnetic field and flux distribution around the coil in the system. The results of this analysis are shown in Figure 3, where the domain is sliced along the x-y plane and magnetic flux density is shown in Figure 3a. Similarly, the magnetic flux density near the coil and sample space is displayed (Figure 3c and Figure 3d), and the magnetic field and magnetic potential on the sample space were analyzed and are summarized (Table 1).

These results serve as a basis for the Case 2 analysis, where the number of coils, turns per coil, and current per turn were chosen arbitrarily and are modular in the design of the solenoid.

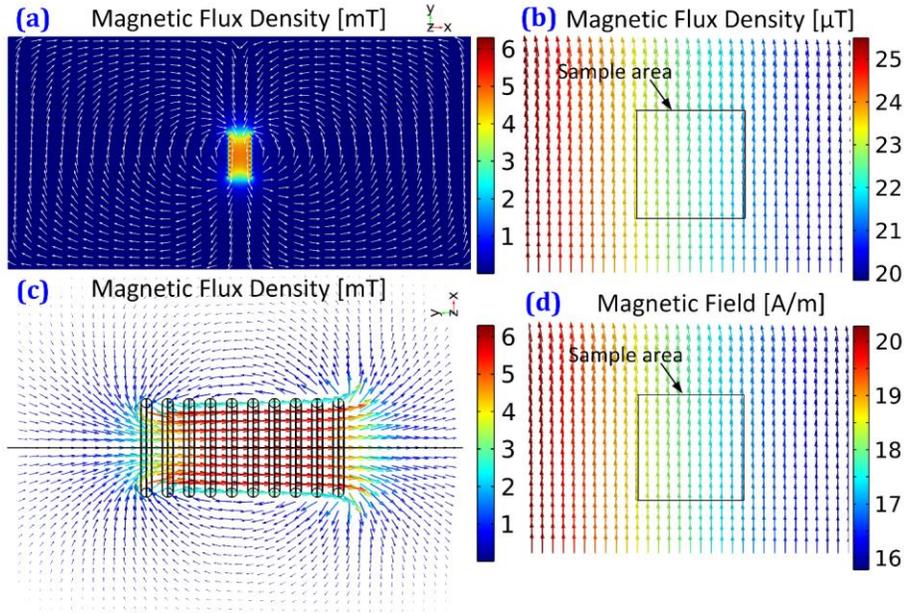

**Figure 3**: CASE1 Magnetic flux density results (a) Cross sectional view of the magnetic field in xy plane. The field lines represents the direction of the flux density and the contour color represents the intensity. (b) Close-up view of the sample space to see the magnetic flux density (c) Close-up view of the coil for the magnetic flux density distribution. (d) Magnetic field near to the sample area.

**Table 1**: Magnetic potential and magnetic field estimated on sample surface

|  | Magnetic Potential | | | Magnetic Field | | |
|---|---|---|---|---|---|---|
|  | Wb/m | | | A/m | | |
| Face | $A_x$ | $A_y$ | $A_z$ | $H_x$ | $H_y$ | $H_z$ |
| $X_{low}$ | -5.33E-09 | 1.31E-08 | 4.10E-07 | -1.51 | 18.53 | 2.62E-04 |
| $X_{high}$ | 5.06E-09 | 1.20E-08 | 3.70E-07 | -1.345 | 17.06 | 2.62E-04 |
| $Y_{low}$ | -2.96E-09 | 6.12E-09 | 3.85E-07 | -0.688 | 17.83 | 2.48E-04 |
| $Y_{high}$ | -4.04E-09 | 1.36E-08 | 3.96E-07 | -2.165 | 17.71 | 2.46E-04 |
| $Z_{low}$ | 9.71E-09 | 6.50E-09 | 4.19E-07 | -1.427 | 17.78 | 0.0236 |
| $Z_{high}$ | -8.39E-09 | 1.85E-08 | 4.31E-07 | -1.427 | 17.78 | -0.0232 |

### 3.2. Case 2: Stress and Strain Estimation on Fibrin Sample

In this case the fibrin sample space and embedded MMPs were modeled with a finer mesh. Figure 4 displays the fibrin cube with different concentrations of MMPs (dilute and non-dilute). The outputs from Case 1 (Table 1) were supplied as boundary conditions and used to estimate the local distribution of the magnetic field and magnetic flux density. The results of magnetic flux density and magnetic field for the dilute and non-dilute concentrations are shown in Figure 4c, Figure 4g, Figure 4d and Figure 4h, respectively.

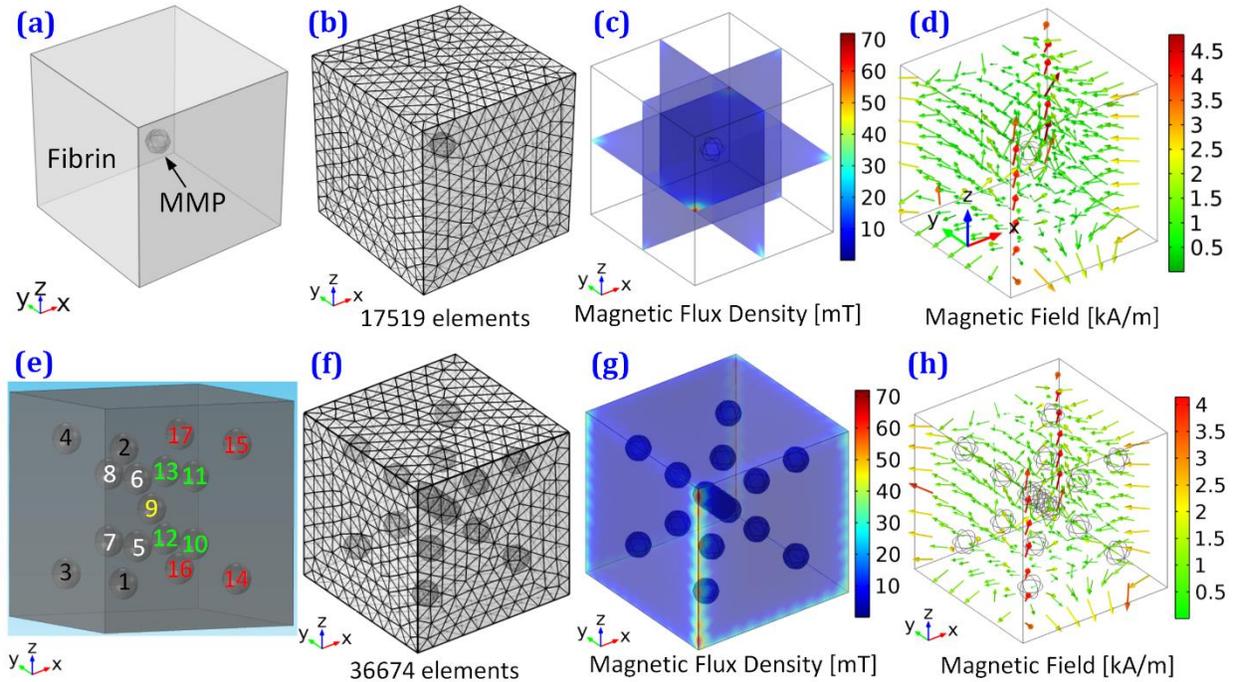

**Figure 4**: Case 2: Magnetic flux density and magnetic field estimation (a,e) 3D model for the fibrin sample with embedded MMPs for dilute and non-dilute cases, respectively. (b,f) Meshed regions of dilute and non-dilute MMP concentrations. (c,g) Magnetic flux density of dilute and non-dilute MMP concentrations. (d,h) Magnetic field orientation for dilute and non-dilute MMP concentrations

The dilute MMP concentration (one MMP/box, 0.0112 g/ml) and non-dilute MMP concentration (17 MMPs/box, 0.191 g/ml) were employed to assess the effects of MMP concentration on electromagnetically induced distortion of the fibrin matrix. The material properties for the MMP and fibrin cube are shown in Table 2 and were derived from the literature [17-21].

**Table 2:** Input material properties for the fibrin/MMP FEA simulations

| Component | Young's modulus | Poisson's ratio | Electrical Conductivity | Relative Permeability | Density |
|---|---|---|---|---|---|
| | MPa | - | S/m | - | kg/m$^3$ |
| MMP | 2.0E5 | 0.291 | 1.0E6 | 2.0E5 | 7,800 |
| Fibrin | 5 | 0.25 | 1.325 | 1.5 | 1,080 |

**Table 3**: Forces acting on MMPs for the non-dilute case

| | MMP 1 | MMP 2 | MMP 3 | MMP 4 | MMP 5 | MMP 6 | MMP 7 | MMP 8 | MMP 9 |
|---|---|---|---|---|---|---|---|---|---|
| $F_x$ [pN] | -17.44 | -7.24 | 20.01 | 8.47 | -2.82 | -2.02 | 3.16 | 2.36 | 0.07 |
| $F_y$ [pN] | -10.98 | 3.80 | 11.38 | -5.22 | 0.15 | -0.79 | 0.02 | 0.70 | -0.01 |
| $F_z$ [pN] | -4.97 | -0.08 | -4.29 | 0.47 | 0.32 | 0.55 | 0.56 | 0.44 | 0.96 |
| | MMP 10 | MMP 11 | MMP 12 | MMP 13 | MMP 14 | MMP 15 | MMP 16 | MMP 17 | |
| $F_x$ [pN] | -2.20 | -1.22 | 2.33 | 1.37 | -15.05 | -3.48 | 14.98 | 4.30 | |
| $F_y$ [pN] | -0.35 | -0.30 | 0.45 | 0.28 | -15.39 | 8.09 | 17.54 | -8.39 | |
| $F_z$ [pN] | 1.00 | 0.69 | 1.16 | 0.74 | 6.00 | -1.47 | 5.73 | 0.03 | |

The estimation of the magnetic field and flux was followed by the estimation of Maxwell's force distribution in the MMPs as per (9) – (11). These forces serve as a coupling parameter between the electromagnetic simulation and the elasticity simulation. For the non-dilute case, the force was estimated as: $F_x$ = 4.42 pN, $F_y$ = -0.017 pN, and $F_z$ = 0.759 pN. The forces acting on the MMPs for the non-dilute MMP case are listed in Table 3. A gravitational force ($F_g$ = -0.879 pN) was applied along the z-axis to every MMP in both cases.

Next, elasticity simulations were performed according to (12) – (13). Transient based FE analysis was Table 4 shows the total displacement for various MMPs.

employed with a 0.1 ms time step of integration for a duration of 100 ms. The lower face ($z_{low}$) was considered as restrained in space to represent the attachment of fibrin to the sample holder and also to avoid free motion induced numerical error. Figure 5a and Figure 5b show the Von Mises stress distribution in the system and Figure 5c shows the variation of the maximum Von Mises stress in the system over time. Figure 5d and 5e show the displacement of the system after reaching steady state for the non-dilute and dilute cases, respectively. Figure 6f shows the time variation of displacement of the system for both non-dilute and dilute MMP cases.

Table 4: Displacements of MMPs for the non-dilute case

|  | MMP 1 | MMP 2 | MMP 3 | MMP 4 | MMP 5 | MMP 6 | MMP 7 | MMP 8 | MMP 9 |
|---|---|---|---|---|---|---|---|---|---|
| u [pm] | **0.447** | **0.556** | 0.192 | 0.189 | 0.105 | 0.137 | 0.097 | 0.163 | 0.134 |
|  | MMP 10 | MMP 11 | MMP 12 | MMP 13 | MMP 14 | MMP 15 | MMP 16 | MMP 17 |  |
| u [pm] | 0.109 | 0.191 | 0.227 | 0.367 | 0.222 | 0.336 | **0.526** | **0.892** |  |

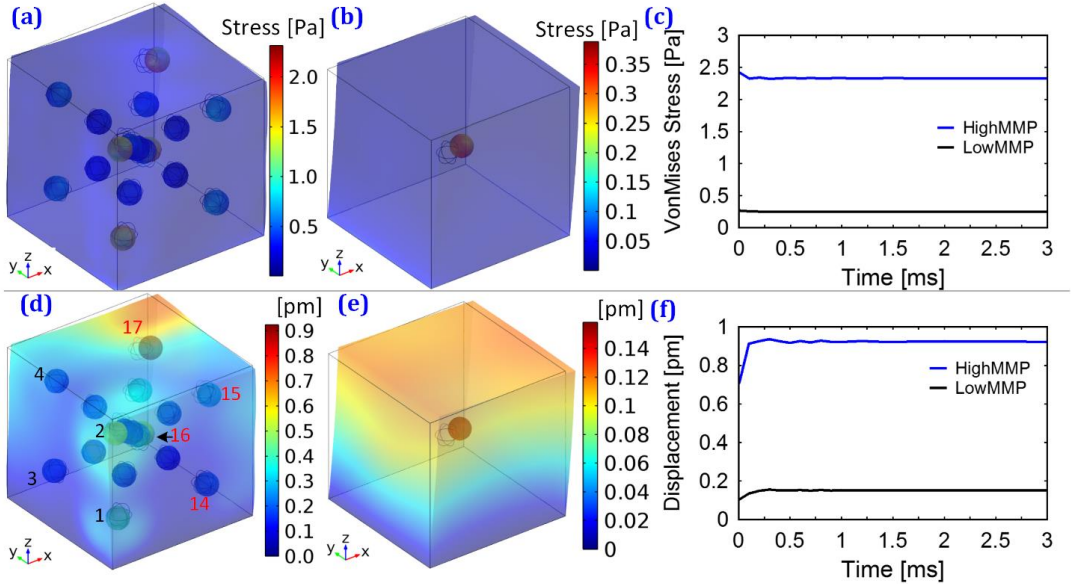

**Figure 5**: Displacement of MMPs due to magnetic and gravitational forces. (a-b) Von Mises stress estimated for the fibrin matrix due to the external loading (magnetic and gravity) for non-dilute and dilute MMP cases. (c) Stress evolution over time. (d-e) Total displacement of the MMPs under stress for both non-dilute and dilute MMP cases. (f) Displacement evolution over time

## 4. Results and Discussion
### 4.1. Case 1: Magnetic Field Estimation for the Entire Domain
The finite element model was employed to represent the fibrin clot sample near a coil. In effect, it produced a magnetic field around the fibrin sample when modeled in the presence of current carrying coils. The coils with the arbitrary configuration of geometry and wire dimensions yielded a magnetic flux density of 25 μT, magnetic field intensity of 18.5 A/m, and a maximum magnetic vector potential of 18.5 nW/m near the sample space. These values were computed and used as input boundary conditions for the sub-model analysis (Case 2).

**4.2. Case 2: Sub-model magnetic field estimation**
The sub model represents the fibrin matrix with embedded MMPs and was meshed with finer tetrahedral elements and applied boundary conditions from Case 1 simulation results. The magnetic vector potential, magnetic field, and magnetic fluxes were estimated on the sample cubic faces by averaging over the surface area. This single valued averaged quantity was employed as the surface boundary condition in the sub model of fibrin.

The Maxwell stress tensor was integrated over the MMP surfaces to obtain the corresponding acting electromagnetic forces. A maximum directional force of 20 pN was observed on MMP #3 which is located on the farthest location from the origin. These forces were computed and supplied as the body forces for the elasticity simulations.

**4.3. Case 2: Elasticity simulations**
The elastic simulations of fibrin with embedded MMPs was conducted by applying the electromagnetic forces and the gravity force on the MMPs. These analyses were initially performed as transient and the results indicate that steady-state was obtained. In addition, the maximum Von Mises stress and total displacement on the MMPs were estimated as 2.33 Pa and 0.92 pm, respectively. These stress values are considerably lower than the yield strength of fibrin, which has been previously reported as 4 kPa in the literature [21]. The aforementioned maximum stress and displacement values demonstrate that the forces engendered by the initial configuration of the coil were not sufficient for the MMPs to experience a high strain.

**5. Extended Parametric Analysis**
To estimate the required force and the necessary coil parameters to achieve plastic deformation in the fibrin matrix, the current study was extended with a parametric sweep analysis. The extended analysis was performed in a sequential manner. First, the forces acting on the MMPs were scaled parametrically to achieve the deformation in the fibrin sample. Upon determination of the force which caused plastic deformation in the fibrin matrix, the next step was to estimate the magnetic field strength which engendered the corresponding magnitude of force. To achieve this, we parametrically altered the applied magnetic field and potential on the surfaces of the fibrin sub model. Finally, the geometric and electrical properties of the coil were designed to create the appropriate magnetic field and magnetic potential.

The variation of displacement and von Mises stress was computed while the forces on the MMPs were adjusted. The results show that the fibrin matrix reaches a yield stress of 4 kPa when the maximum force on the MMPs is 36 nN. The variation of the maximum force on the MMPs (while the magnetic field strength was adjusted) was calculated (Figure 6). The magnetic field strength on the surface (~730 A/m) was applied to achieve the maximum electromagnetic force of 36 nN. FE simulations are corroborated by physiological calculations. From a physiological standpoint, it is known that fibrin clots retract normally in the presence of platelets [22-27]. However, in this study, the forces exerted by MMPs were representative of tensile forces within the elastic regime of fibrin when stretched. Considering the drag force on a fibrin cube to be: $F_D = \frac{1}{2}\rho v^2 A C_D$, where $\rho$ is the density, $v$ is the velocity of blood flow (~1 m/s for normal blood flow), $A$ represents the surface area of the cube ($s^2$) and $C_D = \frac{1}{Re}$. Reynold's number *(Re)* is calculated as: $Re \sim \frac{\rho v s}{\mu}$, where $\mu$ represents the viscosity of the blood (~0.003 Pa.s) and *s* represents the dimension of the side of the cube (20 μm). With this information, the drag force due to normal blood flow in this system is 30 nN, which is lower than the force to achieve plastic deformation in the sample (36 nN total force acting on MMPs). Biologically and clinically, this implies that under normal flow conditions for this system, the drag force alone would not be sufficient to achieve plastic deformation in the sample and thus an external force would be necessary to plastically deform the fibrin clot. In a recent study by Nam et al [28], fibrin matrices were experimentally tested to determine the degree of plasticity in the samples when subjected to a constant stress. At long timescales (3,600 s), the degree of plasticity was measured at a constant value of 0.2 for stress values ranging from 10 Pa to 100 Pa. For shorter timescales (~300 s), the degree of plasticity was much lower (~0.01) for stress values ranging from 10 Pa to 100 Pa, indicating very low irreversible damage in the samples. The shorter timescales are more representative of our case, and again indicate that an external force (such as an electromagnetic force) would be necessary to achieve plastic deformation in the fibrin samples with embedded MMPs.

One minor caveat of the current study was the use of an elastic model to represent the fibrin system. It is well established that fibrin is a viscoelastic material [6, 29-34]; however, the purpose of this study was to ascertain the effects of inclusions (MMPs) within a fibrin matrix when subjected to a magnetic field. Utility of a viscoelastic model in this case would be extremely difficult and our goal was to determine the plastic limit of the fibrin matrix (irreversible limit) with MMPs for the system. In addition, the fibrin matrix was modeled as a continuum. Future studies will deviate from the continuum model in which the fibrin matrix will be modeled as a polymeric fibrin fiber network with embedded inclusions (MMPs). We also believe that close to plastification, fibrin contacting the rear part of the MMP may not be mechanically sufficient to resist the tension forces and the shear stresses that occur on the upper and lower boundaries of the MMP, and thus close to plastification, a contact model between fibrin and MMP will be employed.

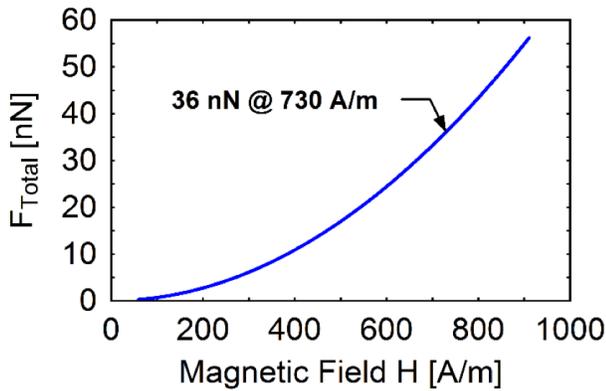

**Figure 6**: Force and magnetic field parametric sweep analysis results. Magnetic field strength was scaled and the total maximum force acting on MMPs was computed.

After estimation of the necessary magnetic field strength of 730 A/m on the fibrin sub model, the next step was to design and calculate the coil parameters ($R, N/c, I$). These parameters represent the coil radius $(R)$, # turns/coil $(N/c)$, and the wire current $(I)$. An independent parametric sweep was performed and it was determined that a configuration with 185 turns/coil with 1 A current/turn (or 19 turns/coil with 10 A of current/turn) is required to plastically deform the fibrin matrix. The coil diameter was chosen as 2 cm for these two studies. In the analysis where the radius of the coil was swept, it was determined that a 4 cm diameter coil was necessary for achieving 400 A/m. Increasing the diameter of the coil would result in a bulky design and hence a fixed diameter of 2 cm was prescribed for the coil. The results show that the required value of magnetic field can be achieved if the total current/coil is maintained above 190 A. Hence for a 10 coil system, a magnetic field can be created to engender plastic deformation in the fibrin clot utilizing the following wire-current relationship.

$$N_{turns} \times I \geq 190 \qquad (14)$$

## 6. Conclusions

In this paper, a mechanical model is presented using electromagnetic forces on a fibrin matrix with embedded magnetic MMPs using finite element analysis. The current carrying coil creates the electromagnetic field and the location of the sample is modeled. The results from the magnetic potential and field on the surfaces of the fibrin sample cube were estimated and performed on a sub domain analysis of the fibrin matrix. The sub domain analysis consisted of both electromagnetic and elasticity simulations. From the elasticity simulations, we estimated the maximum stress induced on the fibrin sample as 2.33 Pa, which was considerably lower than the yield strength of the fibrin clot (4 kPa). This was due to the weak electromagnetic field created by the initial selection of coil parameters, to include current, $N/c$, and $R$. These coil parameters were initially chosen as a result of basic physiological force calculations, which could potentially be applied to an in-vivo fibrin-MMP system.

An extended parametric sweep analysis was performed, and a relationship was found between the coil parameters $N/c$ and $I/N$ which was required to create plastic deformation in the fibrin matrix. Based on these findings studies, a sample coil with a 2 cm coil diameter, 2 A $I/N$ and $100$ $N/c$ can create the sufficient electromagnetic field at a distance of 3 cm away from the coil center to engender plastic deformation conditions. The fibrin matrix reaches the yield stress of 4 kPa when the maximum force on the MMPs is 36 nN. These studies can be used to design in-vitro experiments to assess how fibrin matrices with the inclusion of embedded magnetic microparticles can be perturbed in the presence of an electromagnetic field.

**Acknowledgments**

Research reported in this publication was supported by the National Heart, Lung, and Blood Institute of the National Institutes of Health under Award Number K01HL115486. The content is solely the responsibility of the authors and does not necessarily represent the official views of the National Institutes of Health.